\documentclass[12pt]{article}

\usepackage{amssymb,amsmath}
\usepackage{array}

\makeatletter
\def\alt{\mathrel{\mathpalette\gl@align<}}
\def\agt{\mathrel{\mathpalette\gl@align>}}
\def\gl@align#1#2{\lower.6ex\vbox{\baselineskipZ@skip\lineskipZ@
\ialign{$\m@th#1\hfil##\hfil$\crcr#2\crcr\sim\crcr}}}
\makeatother

\begin{document}
\begin{flushright}
MIFP-09-11 \\
March, 2009
\end{flushright}
\vspace*{1.0cm}

\begin{center}
\baselineskip 20pt
{\Large\bf
Flavor Symmetry in Gauge-Higgs-Matter Unified Orbifold GUTs
}

\vspace{1cm}

{\large
Yukihiro Mimura
}
\vspace{.5cm}

{\baselineskip 20pt
\it 
Department of Physics, Texas A\&M University,
College Station, TX 77843-4242, USA
}
\vspace{.5cm}

\vspace{1.5cm}
{\bf Abstract}
\end{center}

We construct an orbifold model in which
all the standard model particles are unified in
a gauge multiplet in higher dimensional supersymmetric gauge theory.
We find that a flavor symmetry has to remain in four dimensions
due to the discrete charge conservation for orbifold conditions
if the colored Higgs components in the gauge multiplet are projected out.
When the flavor symmetry originates from the $E_8$ bulk gauge symmetry,
the successful unification model can be constructed.
In the model, all the Dirac Yukawa couplings for quarks and leptons
are generated by the higher dimensional gauge interaction.


\thispagestyle{empty}

\newpage

\addtocounter{page}{-1}

\section{Introduction}
\baselineskip 18pt

The standard model (SM) is well established to describe the physics
below weak scale.
Theoretically, it is expected that there exists a model beyond SM.
The conceptual motivation to consider the model beyond SM
is to understand the
variety of particles as well as the parameters in SM.
In fact, the content of particles in SM
is a collection of widely disparate fields:
gauge bosons coming in three factors (color, weak and hypercharge),
three replicated families of chiral fermions
($q$, $u^c$, $d^c$, $\ell$ and $e^c$),
and a scalar Higgs boson to break electroweak symmetry
and give masses to the chiral fermions.
The Higgs scalar has a quadratic divergence
in its mass squared, and thus the electroweak scale
is not stable quantum mechanically if the cutoff scale is very high
such as the Planck scale.
Therefore, there must exist a theory beyond SM around the TeV scale.
Besides, the masses and mixings of the fermions 
are generated by the Yukawa interactions with the Higgs boson,
which are the most of the parameters in SM. 
In such a sense, the nature of the Higgs boson
is a key ingredient to go beyond the standard model.

Supersymmetry (SUSY) is one of the most promising candidates
to construct a model beyond SM.
The quadratic divergence of the Higgs boson mass squared
is canceled due to the unification of couplings of 
fermions and bosons.
Therefore, the electroweak scale is stable once it is generated,
and the model at ultra high energy scale can be extrapolated
from the weak scale physics.
Although the particles get doubled naively in the SUSY extension,
these new SUSY particles around the weak scale
add an additional attraction
in minimally extended SUSY standard model (MSSM).
The measured three gauge couplings can be unified at a scale
through renormalization group running.
That provides us a motivation of grand unified theory (GUT)~\cite{Georgi:1974sy} 
that the gauge symmetry of the SM
is unified in a simple group, such as $SU(5)$ and $SO(10)$.
The quarks and leptons can be unified in a larger 
representation, especially in $SO(10)$ model.
However, then the Higgs sector of the model becomes complicated. 
Naively, the Higgs doublet in SM is also unified in 
a larger representation.
Then, there should be colored Higgs particles,
which may cause a rapid proton decay rate rather
than the current experimental bounds
in SUSY GUT \cite{Sakai:1981pk}.
Therefore, the colored Higgs fields should be heavy,
or their couplings to quarks and leptons needs to be 
forbidden.

The idea of extra dimensions is attractive
to build a model beyond SM \cite{Antoniadis:1990ew}. 
%
The extra dimensions are compactified, and the orbifold boundary conditions
can be used to break symmetries
\cite{Kawamura:1999nj,Hall:2001pg,Mimura:2002te}.
When it is applied in GUT models,
the colored Higgs particles can be projected out,
while Higgs doublet can be light as a four-dimensional (4D) zero mode.
The proton decay operator via the colored Higgs fields
is forbidden in the model~\cite{Kawamura:1999nj}.
Though the gauge symmetry is explicitly broken by the orbifold conditions,
the gauge couplings can be unified when the brane localized gauge interaction
is suppressed by a large volume of the extra dimensions \cite{Hall:2001pg}.
In such fashions, the idea of gauge-Higgs unification~\cite{Manton:1979kb} is revived \cite{Dvali:2001qr}.
The scalar Higgs fields can be unified with
the gauge fields in some higher dimensional vector fields.
In a simple orbifold boundary condition, the gauge symmetry is broken
since the broken generators of gauge bosons
for 4D coordinates are projected out.
At that time, the broken generators for extra dimensional components
can have massless modes, and the Wilson line operator can be identified as the Higgs 
bosons to break the symmetry remaining in 4D \cite{Hosotani:1983xw}.
Interestingly, this idea is compatible to extending the 
SM gauge symmetry to a larger gauge group like GUTs. 
Besides, the mass of the Higgs scalar is forbidden by gauge invariance,
and thus it can remain at low energy
when SUSY is combined in the model.

The interesting consequence of the gauge-Higgs unification
is that the Yukawa interaction can originate from
the gauge interaction when fermions are also 
higher dimensional bulk fields~\cite{Burdman:2002se,Gogoladze:2003bb}.
Actually,
the 4D zero modes of fermions can be chiral in orbifold projections,
and 
the higher dimensional extension of the fermion kinetic term
with covariant derivative,
$\bar\psi \gamma^\mu (\partial_\mu - i g A_\mu) \psi$,
include Yukawa coupling when the gauge fields with higher dimensional
components are identified as Higgs fields.
In the left-right symmetric construction of the model~\cite{Shafi:2002ck},
the matter representation to realize the gauge-Yukawa unification
can be much simpler than that of the SM construction,
and the actual unification of gauge and Yukawa coupling constants
can be realized~\cite{Gogoladze:2003bb}.
In the models
of 5D ${\cal N} =1$ SUSY $S^1/Z_2$ orbifold
with bulk gauge symmetries such as $SO(11)$ and $SU(8)$,
which break down to Pati-Salam (PS) symmetry group
 $G_{\rm PS} = SU(4)_c \times SU(2)_L \times SU(2)_R$ \cite{Pati:1974yy} in 4D,
matter fields are unified in hypermultiplets, and
 all three gauge couplings and third generation
Yukawa couplings (top, bottom, tau and Dirac tau neutrino)
can be unified.
Actually, the prediction of the top quark mass can agree
with the experiment if we take into account the threshold corrections
\cite{Gogoladze:2003pp}.
If we say it inversely,
we will be able to check if the unification of gauge and Yukawa couplings is 
realized in the future since 
the LHC and ILC experiments can provide us more accurate
Yukawa couplings above TeV scale.
It is important that 
the unification of the gauge and Yukawa coupling constants can be 
a signal of extra dimensions at ultra high energy scale.
Therefore, we should investigate models in which gauge and Yukawa 
unification can happen.

In SUSY extensions, 
the matter fermions can be unified in higher dimensional gauge 
multiplets \cite{Chaichian:2001fs,Watari:2002tf,Li:2003ee},
especially when the model consists of $N=4$ vector multiplet in 4D language
such as in 6D ${\cal N}=(1,1)$ SUSY.
Interestingly, three replications of family can be obtained
in the $T^2/Z_3$ orbifold \cite{Watari:2002tf}.

The hypermultiplet which is adjoint representation
under the bulk gauge symmetry in 5D ${\cal N}=1$ SUSY $S^1/Z_2$ orbifold model
can be incorporated into the gauge multiplet in 6D ${\cal N}=(1,1)$
SUSY orbifold models.
It is found that all matter species for one family and Higgs doublets
as well as gauge fields in SM can be unified
in 6D ${\cal N} = (1,1)$ SUSY $SU(8)$ gauge multiplets 
with $T^2/Z_6$ orbifold \cite{Gogoladze:2003ci}.
The three gauge couplings and the third generation Yukawa couplings can be also unified
in the model.
Since no other bulk matter fields can be introduced,
the model can explain why only third family is heavy.
The other gauge groups \cite{Gogoladze:2003yw,Gogoladze:2003cp} 
and other extensions including seven dimensional 
models \cite{Gogoladze:2005az,Gogoladze:2006ps}
have also been considered.

In Ref.\cite{Mimura:2009yq},
we investigated the orbifold models in which 
three families of chiral fermions and Higgs fields, as well as gauge fields
are the zero modes of the bulk gauge multiplet.
When three families are the zero modes of the gauge multiplet, 
the dimension five proton decay operator will be dangerous
if the colored Higgs fields are also the zero modes.
The hierarchy of the Yukawa coupling for fermions
is obtained from the Higgs doublet mixings in such models.
If the colored Higgs fields couple with the first generation quarks and leptons
by the bulk gauge interaction,
the flavor suppression in the proton decay amplitude will not be applied.
Then, the model will be ruled out even if the mass of the colored Higgs
fields are of the order of the Planck scale.
Therefore, one should project out the colored Higgs fields from the 
4D zero modes, which is the original motivation of the orbifold GUTs.
In this paper, 
we study the condition
that the colored Higgs fields are projected out 
and the chiral families of fermions are kept to be zero modes.
We find that a flavor symmetry has to remain in 4D
to satisfy this condition.
The symmetry is non-Abelian global or gauged flavor symmetry
which originates from the $R$ symmetry or bulk gauge symmetry.
%
%

We study 
the phenomenological models with the bulk flavor symmetries. 
In the 6D $SU(8)$ $T^2/Z_3$ orbifold model \cite{Gogoladze:2007nb},
there is a global $SU(3)$ flavor symmetry in the bulk gauge interaction.
As a result of the flavor symmetry,
two eigenvalues are degenerate for the fermion masses from the
bulk Yukawa coupling.
Since the brane localized interaction is considered to be small
by a volume suppression of the extra dimension,
the second generation has to mix with brane localized field
to build a phenomenological model.
Then, the light linear combination of the second generation is almost
the brane fields, 
%
and in that sense, all of the three generations of our fermions
are not the zero modes of the bulk gauge multiplet.
On the other hand, if the flavor symmetry originates from the $E_8$ bulk gauge symmetry,
$SU(4)$ flavor symmetry is possible.
In this case, if we add one family of anti-chiral fermions
as brane-localized fields,
three-family model can be constructed.
In the model, all the fields in SM are the zero modes
of bulk gauge multiplet,
and all the Dirac Yukawa couplings in SM originate
from the bulk gauge interaction.
The Yukawa hierarchy and the flavor mixings for quarks and leptons can be obtained
from the Higgs mixings as well as the 
mixings with the fourth generation of the matter.

This paper is organized as follows:
In section 2, we introduce the
orbifold models which we consider in this paper.
In section 3, we obtain the conditions
to project out the colored Higgs fields
and keep the chiral families to be the 4D zero modes.
The flavor symmetry has to remain in 4D.
In section 4, we discuss the possibility
of the models with the bulk flavor symmetries.
In section 5, we construct a phenomenological model
with $SU(4)$ flavor symmetry,
in which all fields in SM are unified in the higher dimensional 
gauge multiplet.
Section 6 is devoted to the
conclusions.

\section{Gauge, Higgs and matter unification}

In this section, we will briefly study the higher dimensional orbifold model,
which is used to construct a model where
gauge, Higgs and matter fields
are unified in the higher dimensional SUSY gauge 
multiplet \cite{Gogoladze:2003ci,Gogoladze:2003yw,Gogoladze:2003cp}.
We will consider 10D ${\cal N}=1$ SUSY model
to describe the theory generally, but 6D and 8D models can be also considered.
We consider the extra dimensions are compactified over a flat
$T^2/Z_{n_1} \times T^2/Z_{n_2} \times T^2/Z_{n_3}$
orbifold.
The formalism of the higher dimensional models can be
seen in Ref.\cite{Marcus:1983wb}.

From a 4D point of view, the 10D ${\cal N}=1$ gauge multiplet
is recognized as one $N=4$ multiplet which consists of one ${N}=1$
vector superfield $V$ and three chiral superfields $\Sigma_i$ $(i=1,2,3)$.
The scalar components of the chiral superfields $\Sigma_1, \Sigma_2$, and $\Sigma_3$ 
are
$A_5 - i A_6$, $A_7-i A_8$, and $A_9 - i A_{10}$, respectively.
We define the extra dimensional coordinates as $z_1 = x_5+i x_6$, $z_2 = x_7 + i x_8$, and $z_3 = x_9+i x_{10}$.
The orbifold transformations ${\bf R}_i$ are $z_i \to \omega z_i$,
where $\omega = e^{2\pi i/n_i}$.
The transformation ${\bf R}_i$ can also act on the internal symmetry of the Lagrangian.
The internal symmetry in our class of models is the 
product of $R$ symmetry and $Aut(G)$.
This extension of ${\bf R}_i$ can break SUSY as well as the bulk gauge group $G$.
Depending on the discrete charge assignment,
the 4D $N=4$ SUSY can be broken down to $N= 0,1,2$.

If at least $N=1$ SUSY remains in 4D,
the orbifold conditions of the superfields $V$ and $\Phi_i$
are given as
\begin{eqnarray}
V(x^\mu,\bar\omega \bar z_i,\omega z_i) &=& R_i[V(x^\mu,\bar z_i,z_i)], \\
\Sigma_1(x^\mu,\bar\omega \bar z_i,\omega z_i) &=& \bar \omega^{k_i} R_i[ \Sigma_1(x^\mu,\bar z_i,z_i) ], \\
\Sigma_2(x^\mu,\bar\omega \bar z_i,\omega z_i) &=& \bar \omega^{l_i} R_i[ \Sigma_2(x^\mu,\bar z_i,z_i) ], \\
\Sigma_3(x^\mu,\bar\omega \bar z_i,\omega z_i) &=& \bar \omega^{m_i} R_i[ \Sigma_3(x^\mu,\bar z_i,z_i) ],
\end{eqnarray}
where $R_i$ acts on the gauge algebra. 
Since there is a higher dimensional version of trilinear gauge interaction term
in Lagrangian,
$k_i+l_i+m_i  \equiv 0$ (mod $n_i$) needs to be satisfied.
Also, we need $k_1 = 1$, $l_2 =1$ and $m_3 = 1$ to make the lagrangian invariant.
Therefore, one of $(k_i,l_i,m_i)$ has to be 1.
From a geometrical consequence, $n_i$ has to be 2,3,4,6.

By the orbifold conditions, the gauge fields with broken generators
are projected out, while
the chiral superfields with the corresponding broken generators
can have massless modes, which can be identified to the matter and Higgs fields.
Because the scalar components of the superfields $\Sigma_i$ are higher dimensional 
gauge fields,
the bulk gauge interaction includes the term $f_{abc} \,\Sigma_1^c \Sigma_2^b \Sigma_3^c$
in the superpotential in 4D,
where $f_{abc}$ is a structure constant of the gauge group.
Therefore, 
if the matter and Higgs fields are zero modes from the gauge multiplet,
the Yukawa couplings to generate the fermion masses 
can originate from the bulk gauge interaction. 


\section{Projecting out the colored Higgs fields with family unification}

The broken generators for ADE gauge groups, such as $SU(N)$, $SO(2N)$, $E_6$, $E_7$
and $E_8$, often include the matter and Higgs representations
under their subgroups.
For example, when $SU(8)$ bulk gauge group is broken down to
$SU(4)\times SU(2)\times SU(2) \times U(1)^2$,
one set of matter and Higgs representations are included in the broken generators \cite{Gogoladze:2003ci}.
In the case of $E_8 \to E_6 \times SU(3)$,
the broken generator is $({\bf 27},{\bf 3})+c.c.$, which can be identified to three families of matter
and Higgs representations.
The detail of the family unification
is found in \cite{Mimura:2009yq}.


Suppose that the broken generators include the representations for 
two quark doublets $q_1,q_2$,
which can be realized in $SO(16)$ and $E_7$ \cite{Gogoladze:2003yw,Gogoladze:2003cp}.
We also suppose that
$q_1$ component in the adjoint representation is the zero mode in the superfield $\Sigma_1$,
and $q_2$ component is the zero mode in $\Sigma_2$.
Namely, $q_1$ and $q_2$ components in $\Sigma_1$ and $\Sigma_2$ 
have different discrete charges for orbifold.
Then, because of the conservation of discrete charges for the broken generators and 
the relation $k_i+l_i+m_i=0$ for the charge assignments of the superfields,
zero modes of colored Higgs component $h_C$ has to be in $\Sigma_3$
since $q_1 q_2 h_C$ term is included in the cubic coupling of adjoint representation.
When a lepton doublet $\ell_1$ has zero modes in $\Sigma_3$ in addition to the above,
$\ell_1 q_2 \bar h_C$ term is automatically included in the bulk
gauge interaction of the zero modes.
Those colored Higgs couplings are dangerous for a rapid proton decay
once the colored Higgs fields form a Dirac mass term.
Actually, the coupling constant is the same as the gauge couplings,
the usual flavor suppression may not be applied to the decay amplitude,
and it will be excluded even if the colored Higgs mass is of the 
order of the Planck scale.
If more than two quark doublets have zero modes in the gauge multiplet,
we have to care about the dangerous colored Higgs couplings.

Original motivation of the orbifold GUTs is projecting out of the colored Higgs fields
by the orbifold conditions \cite{Kawamura:1999nj}.
In what cases is it possible to project out the colored Higgs fields
keeping the generations of matter fields as zero modes?
Surely, when $h_C$ is projected out by choosing the discrete charges in the above example,
one of $q_1$ and $q_2$ is also projected out due to the discrete charge conservation.

The possibility is the following two cases:

\bigskip

1. The broken generator component $q_1$ is the zero mode in the superfield $\Sigma_1$, and the same component $q_1$ is also the zero mode in $\Sigma_2$.


\bigskip

Since the cubic term of the adjoint representation is tr$\,[\Sigma_1,\Sigma_2] \Sigma_3$,
the symmetric term $q_1 q_1 h_C$ is not included.
Therefore, the zero mode of $h_C$ component is not included in $\Sigma_3$ in this case.
%
%
To realize this situation, 
$\Sigma_1$ and $\Sigma_2$ have to have the same discrete charge $k_i = l_i$.
As a result, the zero modes interaction from the bulk cubic term
has a global non-Abelian flavor symmetry.
The global symmetry originates from the $R$ symmetry in the bulk SUSY.

Three flavor model in this case is already studied in $SU(8)$ 6D $T^2/Z_3$ orbifold model \cite{Gogoladze:2007nb},
where
all $\Sigma_1,\Sigma_2,\Sigma_3$ have the same discrete charges $k_i=l_i=m_i$.
Note that the $SU(8)$ adjoint includes only one generation of matter, and colored Higgs representation
is not included under the broken symmetry $SU(4)\times SU(2)\times SU(2)\times U(1)^2$.
Similar $T^2/Z_3$ orbifold projection can be done even in the case 
where more than two generations are in the adjoint representation
of bulk gauge symmetry, e.g. SO(16).
In those cases, only one of the components of quark doublet has the zero modes, and the generations in 4D 
are replicated by the three chiral superfields. 

\bigskip

2. Both $q_1$ and $q_2$ components are the zero modes in the same superfield $\Sigma_1$.

\bigskip

The $q_1$ and $q_2$ components have the same discrete charges in this case.
Since the bulk interaction is $\Sigma_1 \Sigma_2 \Sigma_3$, 
the colored Higgs component can be projected out keeping the generation
unless the set of $q_1$ and $q_2$ components are also in $\Sigma_2$.
In this case, the generation  of fermions comes from the gauge part,
and the gauged non-Abelian flavor symmetry remains in 4D.

\bigskip

Due to the conservation of discrete charge,
only the above cases are allowed. 
As a result,
there must be a non-Abelian flavor symmetry
when more than two generations of quark doublets
are included in the zero modes of the bulk multiplets
and the colored Higgs component $h_C$ is projected out.

The 4D gauge symmetry is further restricted in order to project out the colored Higgs components.
Suppose that the 4D gauge symmetry is the SM gauge symmetry,
and more than two generation of quark doublets $q_i$ and the lepton doublets $\ell_j$
have zero modes in the superfields.
Then, $q_i$ and $\ell_j$ have to be slotted in different superfields,
and 
$q \ell \bar h_C$ term is included in the bulk interaction of the zero modes.
If $\bar h_C$ is projected out in that situation,
$q_i$ or $\ell_j$ has to be projected out
as long as the bulk gauge symmetry is broken down to SM gauge symmetry.
When both $q_i$ and $\ell_j$ components have zero modes in the same
chiral superfields, the colored Higgs term is not included in the bulk interaction of the zero modes.
At that time, at least, $SU(4)_c [\supset SU(3)_c \times U(1)_{B-L}$] gauge symmetry must remain in 4D
since ($q,\ell$) forms a $SU(4)_c$ multiplet.
Similarly, due to the terms of $u^c d^c \bar h_C + e^c u^c h_C + d^c \nu^c h_C$,
when the colored Higgs fields are projected out completely
and the 4D symmetry is SM gauge symmetry,
only part of the matter species (e.g. $q$ and $u^c$) can be the zero modes.
In other words,
if all of the matter species (including right-handed fermions, $u^c, d^c,e^c,\nu^c$)
with generations are the zero modes of the bulk multiplet, 
and the colored Higgs components (both $h_C$ and $\bar h_C$)
are projected out,
at least $G_{\rm PS} = SU(4)_c \times SU(2)_L \times SU(2)_R$
gauge symmetry must remain in 4D.

If one consider a gauge symmetry which contains $G_{\rm PS}$
(e.g. $SU(6) \times SU(2)_L$ or $SO(10)$),
the colored Higgs components is unified to the matter representations
or the Higgs doublet component.
Therefore, the Pati-Salam symmetry $G_{\rm PS}$ is the basic
building block for the 4D gauge symmetry to project out the colored Higgs fields.

\section{Models with the bulk flavor symmetries}

In the previous section,
we have found that gauged or global flavor symmetry remains
in 4D for the bulk Yukawa interaction
when the generations of matter contents are the zero modes
of the gauge multiplet.
If both left- and right-handed matter representations
are the zero modes in the 4D gauge symmetry $G_{\rm PS}$,
the Higgs bidoublet is also the zero modes
due to the discrete charge conservation.
Then, the Yukawa couplings to generate the fermion masses
originate from the bulk gauge interaction.
However, there is a problem in this scenario.
When there is a non-Abelian flavor symmetry,
two eigenvalues of the fermion masses are degenerate due to the anti-symmetricity
of the bulk Yukawa interaction.
We consider that the brane-localized couplings are suppressed
due to the large volume suppression of extra dimensions.
Then, the bulk Yukawa couplings originated from the gauge interaction
dominates the 4D Yukawa couplings.
Therefore, the degeneracy of the fermion masses
cannot be solved by the brane interaction of the zero modes,
and the model 
is not phenomenologically viable.

In order to solve the degeneracy of the fermion masses,
brane-localized vector-like matter fields
are introduced \cite{Gogoladze:2007nb}.
In this resolution, however,
the second generation matter fields
are almost switched with the brane-localized fields.
Therefore, the Yukawa coupling for the second generation
does not originate from the bulk gauge interaction,
but from the brane-localized interaction.
In the sense, this resolution does not provide a
complete unification scenario of the gauge, Higgs and matter fields
as well as the Yukawa couplings,
although the three chiral matter fields can be zero modes
of the bulk gauge multiplet.
Actually, the second generation of matter has to be always replaced
with the brane fields
as long as the number of the matter replication
is three and the Yukawa matrix is anti-symmetric.

If there are four times replication of bulk matter fields,
the degeneracy can be resolved 
without replacing them with the brane fields.
Consider $SU(4)$ flavor symmetry.
In this case, the eigenvalues are also degenerate due to the anti-symmetricity
of the bulk interaction.
%
When we introduce one generation of anti-chiral matter brane fields,
one linear combination of the four generations become massive by forming a Dirac mass.
At that time, since the brane field is anti-chiral, it is not mixed with
the bulk matter fields.
Due to the mixing with the brane field (breaking of the $SU(4)$ symmetry),
the anti-symmetricity of the fermion mass matrix is resolved, and the hierarchical pattern
can be obtained.
In this situation, the complete unification scenario can be realized :
%
All the particles in SM are the zero modes of the bulk fields 
and all the Dirac Yukawa couplings originate from the higher dimensional gauge interaction.
Effective Yukawa couplings for fermions in the standard models are the 
gauge coupling multiplied by the mixings with the heavy 4th generation.

If the flavor symmetry is global originating from the $R$ symmetry,
the maximal symmetry is $SU(3)$ because there are only three chiral
superfields.
Therefore, to realize the flavor $SU(4)$ symmetry,
it has to originate from the bulk gauge symmetry.
Only possibility of the bulk gauge symmetry which include
$G_{\rm PS}$ and $SU(4)$ flavor symmetry is $E_8$:
\begin{equation}
E_8 \to SO(10)\times SU(4)_F \to SU(4)_c \times SU(2)_L \times SU(2)_R \times SU(4)_F.
\end{equation}
The corresponding breaking of $E_8$
can be done by $T^2/Z_4 \times T^2/Z_4^\prime$ orbifold.

The discrete charge assignments for the $G_{\rm PS}$ decomposition 
of the $E_8$ adjoint representation are the following \cite{Mimura:2009yq}:
\begin{eqnarray}
&&[0] : ({\bf 15},{\bf 1},{\bf 1}),\ ({\bf 1},{\bf 3},{\bf 1}),\ ({\bf 1},{\bf 1},{\bf 3}),
\qquad [e] : ({\bf 6},{\bf 2},{\bf 2}),
\qquad [x_i-x_j]: A_i^j,
\\ 
&&[x_i] : L_i, \qquad [e+x_i] : \bar R_i,
\qquad [-x_i] : \bar L^i, \qquad [e-x_i] : R^i,
\\
&& [x_i+x_j] : C_{ij}, \quad
[e+x_i+x_j] : H_{ij},
\end{eqnarray}
where
$L_i$ and $\bar R_i$ are the matter representations,
$H_{ij}$ is the Higgs bidoublet,
and $C_{ij}$ is the $SU(4)_c$ sextet which contains the colored Higgs representations,
\begin{equation}
L_i : ({\bf 4},{\bf 2},{\bf 1},{\bf 4}),
\quad
\bar R_i : (\bar {\bf 4},{\bf 1},{\bf 2},{\bf 4}),
\quad
H_{ij} : ({\bf 1},{\bf 2},{\bf 2},{\bf 6}),
\quad 
C_{ij} : ({\bf 6},{\bf 1},{\bf 1},{\bf 6}),
\end{equation}
$\bar L$ and $R$ are the anti-chiral matter representations,
and $A_i^j$ is the adjoint of $SU(4)_F$.
Due to the algebra, 
the followings have to be satisfied:
$x_1+x_2+x_3+x_4 \equiv 0$ and $2e\equiv 0$.

When we choose the discrete charges
as
\begin{equation}
(x_1,x_2,x_3,x_4,e)_{Z_4} = (1,1,1,1,2), \quad (x_1,x_2,x_3,x_4,e)_{Z_4^\prime} = (1,1,1,1,0),
\end{equation}
and the assignments for chiral superfields as
\begin{equation}
(k_1, l_1, m_1 )= (1,3,0), \quad (k_2,l_2,m_2)= (1,1,2),
\end{equation}
the 4D gauge symmetry is $G_{\rm PS} \times SU(4)_F$
and $L_i$, $\bar R_i$, and $H_{ij}$ have zero modes in 
$\Sigma_1$, $\Sigma_2$, and $\Sigma_3$, respectively.
The bulk gauge interaction includes the Yukawa terms
$\epsilon^{ijkl} L_i \bar R_j H_{kl} = L_i \bar R_j H^{ij}$.

We comment briefly on the brane-localized gauge anomalies \cite{Scrucca:2004jn}.
The 4D zero modes cause the gauge anomaly for $SU(4)_F$ symmetry.
In order to cancel the anomaly, we have to introduce brane-localized fields at each 4D fixed point.
When there is 16 anti-fundamental representations or 2 anti-decuplets, the $SU(4)_F$ anomaly
is cancelled.
The anti-decuplets can be used to generate Majorana neutrino masses.
The $SU(4)_F$ sextet do not generate the anomaly.

\section{Building a phenomenological model}

In this section, we will construct a model
in which all Dirac Yukawa interaction originates from the bulk gauge interaction
using the $G_{\rm PS} \times SU(4)_F$ symmetry obtained in the previous section.

The Higgs bidoublet field $H^{ij}$ is sextet under the flavor symmetry $SU(4)_F$.
When we introduce brane fields $h_i : ({\bf 1},{\bf 2},{\bf 2},{\bf 4})$
and $h : ({\bf 1},{\bf 2},{\bf 2},{\bf 1})$,
only one of linear combinations of bidoublets can become light ($6-4-1=1$).
The required brane-localized mass terms are
\begin{eqnarray}
s_{ij} H^{ij} h + s_i H^{ij} h_j,
\label{brane-1}
\end{eqnarray}
where $s_{ij}$ and $s_i$ are $SU(4)_F$ sextet and fundamental representations, respectively.
When those SM singlets acquire vacuum expectation values (VEVs), 
the $SU(4)_F$ symmetry is completely broken.
The VEV $\langle s_{ij} \rangle$ breaks $SU(4)_F$ down to $SU(2)\times SU(2)$,
and $\langle s_{i} \rangle$ breaks $SU(2)\times SU(2)$ to nothing.

We introduce anti-chiral brane fields $\bar L_b: (\bar {\bf 4},{\bf 2},{\bf 1},{\bf 1})$,
and $R_b: ({\bf 4},{\bf 1},{\bf 2},{\bf 1})$,
and brane terms:
\begin{equation}
L_i \bar L_b s^i + \bar R_i R_b s^i,
\label{brane-2}
\end{equation}
where $s^i$ is an anti-fundamental representation under $SU(4)_F$.
We introduce multiple $s^i$'s, and the fields $s^i$ which couples to the right-handed matter $\bar R_i$
can have a $SU(2)_R$ breaking VEVs.
Then, the fermion mass term is
  expressed as
\begin{equation}
(L_1, L_2, L_3, L_4, R_b) 
\left(\begin{array}{cc}
m_{ij} & \begin{array}{c} 0\\ 0 \\ 0 \\ M_5 \end{array} \\
\begin{array}{cccc} M_1 &M_2 &M_3 &M_4 \end{array} & 0
\end{array}
\right)
\left(\begin{array}{c}
\bar R_1 \\
\bar R_2 \\
\bar R_3 \\
\bar R_4 \\
\bar L_b
\end{array}
\right),
\end{equation}
\begin{equation}
m_{ij} = \left(\begin{array}{cccc}
0 &0 &0 &c  \\
0 & 0 & b & 0  \\
0& -b& 0& a  \\
-c& 0& -a& 0  
\end{array}
\right),
\end{equation}
where $M_i$ comes from the VEVs of $s^i$, and $a$, $b$ and $c$ are weak scale VEVs of Higgs bidoublets
(multiplied by 4D gauge coupling):
e.g. $a^2+b^2+c^2 = g^2 v_u^2$ for up-type quark mass matrix,
where $v_u$ is the VEV of the up-type Higgs field $H_u$.
By $SU(4)_F$ rotation, 
we can choose that (1,5),(2,5),(3,5) elements in the mass matrix are zero
keeping $m_{ij}$ to be anti-symmetric. 
%
By the remaining $SU(3)_F$ rotation, (1,2),(1,3),(2,4) components can be made zero.
Note that only one linear combination of the Higgs doublets is light,
and the ratios $c/a$ and $b/a$ are related to the Higgs mixings.

The scale of $M_i$ relates to the $SU(4)_F$ breaking,
and one of linear combinations of four matter fields becomes massive,
and three generations of matter remain light.
Using a unitary matrix $U$ where $(M_1,M_2,M_3,M_4) U = (0,0,0,M)$,
the light fermion's mass matrix is obtained as ${\cal M}_{ij} = (m U)_{ij}$ ($i,j = 1,2,3$).
The unitary matrix can be parameterized
as
\begin{equation}
U = \left( \begin{array}{cccc}
		\cos \theta_1 & -\sin \theta_1 & 0 &0 \\
		\sin\theta_1 & \cos \theta_1 & 0 & 0 \\
		0&0&1&0\\
		0&0&0&1
	   \end{array}
    \right)
 \left( \begin{array}{cccc}
		1 & 0 & 0& 0\\
		0&\cos \theta_2 & -\sin \theta_2  &0 \\
		0&\sin\theta_2 & \cos \theta_2  & 0 \\
		0&0&0&1
	   \end{array}
    \right)
 \left( \begin{array}{cccc}
		1&0&0&0\\
		0&1&0&0\\
		0&0&\cos \theta_3 & -\sin \theta_3 \\
		0&0&\sin\theta_3 & \cos \theta_3  
	   \end{array}
    \right),
\end{equation}
where $\tan\theta_1= M_1/M_2$, $\tan\theta_2 = \sqrt{M_1^2+M_2^2}/{M_3}$,
and $\tan\theta_3 = \sqrt{M_1^2+M_2^2+M_3^2}/{M_4}$.
The $3\times3$ fermion matrix ${\cal M}_{ij}$ can be calculated as
\begin{equation}
{\cal M} = (mU)_{3\times 3} = \left(
\begin{array}{ccc}
	0 & 0 & c \sin\theta_3 \\
	0 & b \sin\theta_2 & b \cos\theta_2 \cos\theta_3 \\
	-b\sin\theta_1 & -b\cos\theta_1 \cos\theta_2 &
		b \cos\theta_1 \sin\theta_2 \cos\theta_3 + a \sin\theta_3
\end{array}
\right).
\label{mass-matrix}
\end{equation}

If $SU(2)_R$ breaking VEVs are not inserted in the Higgs bidoublet mass terms in Eq.(\ref{brane-1}),
$a,b,c$ are common for up- and down-type quarks (up to $\tan\beta$ factor),
but the mixing angles $\theta_{1,2,3}$ can be different 
between up- and down-type quarks by the $SU(2)_R$ breaking insertion in Eq.(\ref{brane-2}),
and the quark's generation mixings can be generated. 
The dimensionful parameters $a$ and $c$ can be made real without loss of generality,
and the phase of $b$ can introduce the Kobayashi-Maskawa phase.
Therefore, in this case, there are totally 9 parameters (+ 1 complex phase) in the light quark mass
matrices in the model.
In general, the $SU(2)_R$ breaking VEVs can be inserted in the Higgs bidoublet mass terms
as well as the matter mass terms,
and so there can be more parameters.

We note that
if only one $s^i$ couples to the matter fields, $M_{1,2,3}$ entries become zero in the above basis.
Then, the anti-symmetricity of the light fermion mass matrix is kept and 
the mass degeneracy is not resolved.
When more than two $s^i$'s couples to the matter fields (and left-right parity is broken
in the brane-localized terms),
the degeneracy can be resolved.




The mass eigenvalues $m_{1,2,3}$ have relations:
\begin{eqnarray}
m_1 m_2 m_3 &\!\!=&\!\! -b^2 c \sin\theta_1 \sin\theta_2 \sin\theta_3, \\
m_1^2+m_2^2+m_3^2 &\!\!=&\!\!
(a^2+c^2)\sin^2\theta_3
+ b^2(1+\cos\theta_3^2+\sin^2\theta_1\sin^2\theta_2\sin^2\theta_3) \\
&&+ ab \cos\theta_1\sin\theta_2\sin2\theta_3, \nonumber \\
m_1^2 m_2^2 + m_2^2 m_3^2 + m_1^2 m_3^2 &\!\!=&\!\!
b^2\left[ (a^2 \sin^2\theta_2+ c^2(1+\sin^2\theta_1 \sin^2\theta_2))\sin^2\theta_3
\right. \\
&&\!\!\left.
+ b^2(\cos^2\theta_3+\sin^2\theta_1 \sin^2\theta_2\sin^2\theta_3)
+ ab \cos\theta_1 \sin\theta_2 \sin2\theta_3
\right]. \nonumber
\end{eqnarray}
There are typically three cases to obtain hierarchical eigenvalues:
\begin{enumerate}
	\item $b,c \ll a$.
	\item $c \ll a\sim b$.
	\item $b \ll a\sim c$.
\end{enumerate}

In the case 1, the eigenvalue of the third generation
is $m_3 \simeq a \sin\theta_3$,
and thus the Yukawa coupling for the third generation can be 
written as $y_3 \simeq g \sin\theta_3$.
In this case, the third generation Yukawa coupling is always smaller than (or equal to)
the 4D gauge coupling $g$.
Therefore, $\tan\beta$ is naively a function of $\sin\theta_3$ for down-type quark sector.

In the case 2, it is necessary that $\cos\theta_3$ and $\sin\theta_2$ are small
to obtain the hierarchy between second and third generation.
The eigenvalue of the third generation is then $m_3 \simeq \sqrt{a^2+b^2}$
and the third generation Yukawa coupling is unified to the gauge coupling $y_3 \simeq g$.
Since bottom Yukawa coupling is large, $\tan\beta$ has to be large $\sim 50$.

In the case 3, the third generation Yukawa coupling is $y_3 \simeq g \sin\theta_3$
similarly to the case 1.
The second generation mass eigenvalue is
$m_2 \simeq b \sqrt{(a^2\sin^2\theta_2+c^2)/(a^2+c^2)}$.
The factor $\sin\theta_1 \sin\theta_2$ has to be small to obtain a small mass for the first generation.
Therefore, if the ratio of $a,b$ and $c$ are same for up- and down-type quarks 
(which can be realized when the insertion of $SU(2)_R$ breaking VEVs in the Higgs doublet mass term is small),
the Yukawa couplings for charm and strange quarks
are almost same (especially when $a \sin\theta_2 \ll c$).
As a result, $\tan\beta$ has to be related to the charm-strange quark mass ratio
($\tan\beta \sim 5-15$).

The quark mixings can be generated especially in the case 2.
The detail of the mixings is noted in Appendix.
The naive relations $V_{cb} \sim m_s/m_b$ and $V_{ub} \sim V_{us} V_{cb}$
are naturally satisfied when $\sin\theta_2$ is small.
The Cabibbo mixing angle is naively $\sim (c/b) \tan\theta_3/\cos\theta_2$.
Therefore, the Cabibbo angle become relatively large when $\cos\theta_3 \to 0$,
which is related to the condition that
the third generation Yukawa coupling constant is the same as the gauge coupling constant,
and the second generation Yukawa coupling is relative small.


The suitable choice of parameters can be obtained when $M_1,M_2,M_4 \ll M_3$.
In that case, one can interpret that $M_3$ and $M_5$ break the $SU(4)_F$ symmetry down to $SU(2)$
symmetry, and the other quantities $M_1,M_2$ break the remaining $SU(2)$ symmetry 
to obtain the fermion hierarchy.
The CP symmetry in the quark mass term is spontaneously broken 
and the Kobayashi-Maskawa phase is obtained from the VEVs of $s^i$.

All the Dirac Yukawa couplings for fermions originate from the gauge interaction,
but the Majorana couplings for neutrinos are introduced by brane-localized interaction.
Thus, there are many numbers of parameters when $G_{\rm PS}$ symmetry is broken,
and the neutrino mixings are less constrained compared to the quark mixings.
To generate the Majorana masses for neutrinos, we need a $SU(4)_F$ anti-decuplet
or (at least) four anti-fundamental representations which couple to the matter fields.

We comment briefly on the $SU(4)_F$ symmetry breaking scale.
The exchange of the $SU(4)_F$ gauge bosons gives rise to flavor changing neutral currents,
and thus the breaking scale has to be much larger than the weak scale
in order to suppress unwanted effects \cite{Maehara:1979kf}.
The Lagrangian has a Peccei-Quinn-like symmetry,
and the $SU(4)_F$ breaking scale (VEVs of $s^i$) is related to the
Peccei-Quinn-like symmetry breaking.
The Peccei-Quinn symmetry breaking is also related to the Majorana mass scale for the right-handed
neutrinos.
The strong CP problem can be solved in this framework
when the Peccei-Quinn symmetry breaking scale is $10^{12}$ GeV \cite{Peccei:1977hh},
similar to the models with $SU(3)$ horizontal symmetry \cite{Kitano:2000xk}.
To generate fermion mass hierarchy, the $SU(4)_F$ will not break at a single scale,
and thus the Peccei-Quinn and $SU(4)_F$ symmetry breaking scales may have a difference
relating to the fermion hierarchy.
At the mass scale of the vector-like fourth generation,
the Peccei-Quinn-like symmetry can still remain,
and thus
the mass of the fourth generation
is expected to be larger 
than the axion scale.

%
%



\section{Conclusion}

We studied the higher dimensional models in which gauge, Higgs and families of chiral fermions
are unified in a SUSY gauge multiplet.
The Yukawa couplings to generate the fermion masses originate from the bulk gauge interaction.

When the first generation of fermions are the zero modes of the bulk fields
and the zero modes of the colored Higgs fields couple to them by the bulk gauge interaction,
it causes the rapid proton decay.
Thus, we should project out the colored Higgs components
in the bulk gauge multiplet.
When the colored Higgs fields are completely projected out by the choice of the
discrete charge assignments for the orbifold condition
and the families of the chiral fermions are the zero modes from
the bulk gauge multiplet,
there must remain a non-Abelian flavor symmetry in 4D.
The flavor symmetry originates from the $R$ symmetry or the bulk gauge symmetry.

When there is a flavor symmetry in the bulk Yukawa interaction,
the fermion mass eigenvalues are degenerate.
Since we expect that the brane interactions are suppressed,
the degeneracy should be resolved by introducing the brane-localized fields.
When there is a $SU(3)$ flavor symmetry
and three chiral families are the zero modes of the bulk gauge multiplet,
one of the families are (almost) switched with the brane fields to
resolve the mass degeneracy.
In this case, there is a merit that the second and third generation hierarchy
is obtained by the volume suppression of the extra dimensions,
and the lightness of the first generation is obtained by
the exponential suppression of the Wilson line operators \cite{Gogoladze:2007nb}.
It is conceptually interesting that
 the replication of three families are caused by the three chiral superfields
in the gauge multiplet.
When there is a $SU(4)$ flavor symmetry
and four chiral families are the zero modes,
the mass degeneracy can be resolved by introducing one
anti-chiral family.
In this case, 
all three generations of the quarks and leptons, Higgs fields and the gauge bosons contained in SM
can be successfully unified in one gauge multiplet.
All the Dirac Yukawa couplings originate from the bulk gauge interaction,
and the effective Yukawa couplings in SM
to generate the hierarchical fermion masses and the generation mixings
are obtained from the mixings of the Higgs doublets 
as well as the mixings with the heavy fourth generation.
The $SU(4)$ flavor symmetry has to be a gauged symmetry
which originates from the bulk $E_8$ gauge symmetry,
and the model is more restricted rather than the case of $SU(3)$ flavor symmetry.

In conclusion, 
a flavor symmetry has to remain in 4D in the family unified orbifold GUTs
if the colored Higgs components are projected out from the 4D zero modes.
%
If the flavor symmetry is $SU(3)$, the fermion hierarchy can be explained by
the suppressions from extra dimensions.
If the flavor symmetry is $SU(4)$ originating from the bulk $E_8$ gauge symmetry,
the hierarchy is obtained by the $SU(4)$ breaking patterns,
and all the Dirac Yukawa couplings to generate fermion masses are introduced by the 
higher dimensional gauge interactions.

\appendix

\section*{Appendix: Note on the diagonalization unitary matrix }

We obtain the diagonalization unitary matrix $V$ ($V M M^\dagger V^\dagger = M^2_{\rm diag}$)
when the mass matrix $M$ is written in the form
of
\begin{equation}
M = \left( 
	\begin{array}{ccc}
	0& 0& qy  \\
	0& w& py  \\
	u& v& x
	\end{array}
    \right),
\end{equation}
as in Eq.(\ref{mass-matrix}).
Since the mass matrix can be always made to be in the above form by an unphysical unitary matrix
from the right-hand side, the expression below will be useful in general.
We define $p^2+q^2=1$ to make the expression simple.
We assume $y,w \ll x,u,v$ and $q \ll 1$.
Without loss of generality, only $w$ has a complex phase. 

When $w = 0$, one can obtain the exact diagonalization matrix as
\begin{equation}
V = \left( 
	\begin{array}{ccc}
	1& 0& 0 \\
	0& c_{23}& -s_{23} \\
	0& s_{23}& c_{23}
	\end{array}
    \right)
    \left( 
	\begin{array}{ccc}
	p& -q& 0 \\
	q& p& 0 \\
	0& 0& 1
	\end{array}
    \right),
\label{diagonalization}
\end{equation}
where $s_{23} = \sin\theta_{23}$, $c_{23} = \cos\theta_{23}$ and
$\displaystyle \tan 2\theta_{23} = \frac{2xy}{x^2+u^2+v^2-y^2}$.
The 13 element of $V$ is exactly zero.
Suppose that the above mass matrix is for the down-type quarks and the
up-quark mass matrix is already diagonalized in the basis.
Then the CKM (Cabibbo-Kobayashi-Maskawa) quark mixing matrix is
$V_{\rm CKM} = V^{\dagger}$,
and we obtain $V_{ub} = q s_{23}$.
Therefore, the limit $w\to 0$ is suitable to understand the
empirical relation $V_{ub} \sim V_{us} V_{cb}$.
The strange-bottom quark mass ratio is obtained
as $m_s/m_b \simeq V_{cb} v/x$ in the limit,
and thus the suitable relation between the ratio and $V_{cb}$ 
can be obtained when $v\sim x$.

When $w\neq 0$, the diagonalization matrix is obtained approximately as
\begin{eqnarray}
V \simeq 
     \left( 
	\begin{array}{ccc}
	c_{12}& -s_{12}& 0 \\
	s_{12}& c_{12}& 0 \\
	0& 0& 1
	\end{array}
    \right)
       \left( 
	\begin{array}{ccc}
	c_{13}& 0& -s_{13} \\
	0& 1& 0 \\
	s_{13}& 0& c_{13}
	\end{array}
    \right)
    \left( 
	\begin{array}{ccc}
	1& 0& 0 \\
	0& c_{23}& -s_{23} \\
	0& s_{23}& c_{23}
	\end{array}
    \right)
    \left( 
	\begin{array}{ccc}
	p& -q& 0 \\
	q& p& 0 \\
	0& 0& 1
	\end{array}
    \right),
\end{eqnarray}
where 
$\displaystyle \tan 2\theta_{23} = \frac{2(xy+pvw)}{x^2+u^2+v^2-y^2}$, and
\begin{equation}
s_{13} \simeq - q w (c_{23} v + s_{23} p w)/m_3^2,
\qquad
s_{12} \simeq q w (s_{23} v - c_{23} p w)/m_2^2,
\end{equation}
%
%
%
and
$m_2^2 \simeq y^2+w^2 - \frac14 \tan^2 2\theta_{23} m_3^2$, 
$m_3^2 \simeq u^2+v^2+x^2$.
%
%
%
%
Then, we obtain
\begin{equation}
V_{21} \simeq q + p s_{12}, \quad
V_{31} \simeq q s_{23} + p s_{13},
\quad
V_{32} \simeq p s_{23}.
\end{equation}
%
%
%
Here we do not consider the phase of $w$ to make the expression simple,
but the extension including the complex phase can be easily done.
We note that when the correction from $w$ is small, the
Kobayashi-Maskawa phase cannot be large since only $w$ can have a physical complex phase.
Actually, the experimental measurement gives $V_{ub}/(V_{us}V_{cb}) \simeq 0.5$,
and there has to be a correction from $w$.
When $w$ is large enough to modify the relation, the proper range of down quark mass
can be obtained.

When $u,v \ll x$ and $m_2 \simeq w$, $s_{12}$ is comparable to $q$ 
and $V_{12}$ is cancelled to be small.
At that time, the expression above is not a good approximation.
In that case, $V_{13} \simeq qy/x$ and 
$V_{12} \simeq -qvy/(wx)$.

%

We note that the neutrino mixings are large if $u\sim v \sim x$ 
and the charged-lepton
mass matrix is transpose of the down-type quark mass matrix,
which can be naively realized in $SU(5)$ GUTs.
If the neutrino mass matrix is diagonalized in this basis,
the neutrino mixing matrix is $V^*$ (instead of $V^\dagger$)
and thus the
13 element of the mixing matrix ($U_{e3}$) is naturally small
in the limit $w\to 0$.
Therefore, the basis is useful to see the mixings of quarks and leptons.

\section*{Acknowledgments}

We thank Prof. S. Nandi for useful discussions.
This work 
is supported in part by the DOE grant
DE-FG02-95ER40917.

\end{document}